\documentclass[epj]{svjour}
\usepackage{graphics}
\usepackage{amsmath}
\usepackage{amssymb}
\usepackage{epsfig}

\newcommand{\reaktion}{\mbox{$\vec{p}p\rightarrow\,pp\:\!\eta$ }}

\def\vec{\mathaccent"017E}

\title{First close-to-threshold measurement of the analyzing power
  $\boldsymbol{A_y}$ in the reaction $\boldsymbol{\reaktion}$}
\titlerunning{First close-to-threshold measurement of the analyzing power  $A_y$ in the reaction $\reaktion$}
\author{ 
         P.~Winter\inst{1}              \and
         H.-H.~Adam\inst{2}             \and
         A.~Budzanowski\inst{3}         \and
         R.~Czy{\.{z}}ykiewicz\inst{4}  \and
         D.~Grzonka\inst{1}             \and
         M.~Janusz\inst{4}              \and
         L.~Jarczyk\inst{4}             \and
         B.~Kamys\inst{4}               \and
         A.~Khoukaz\inst{2}             \and
         K.~Kilian\inst{1}              \and
         P.~Kowina\inst{1,6}            \and
         T.~Lister\inst{2}              \and
         P.~Moskal\inst{1,4}            \and
         W.~Oelert\inst{1}              \and
	 C. Piskor-Ignatowicz\inst{4}  \and
         T.~Ro{\.{z}}ek\inst{1,6}       \and
         R.~Santo\inst{2}               \and
         G.~Schepers\inst{1}            \and
         T.~Sefzick\inst{1}             \and
         M.~Siemaszko\inst{6}           \and
         J.~Smyrski\inst{4}             \and
         S.~Steltenkamp\inst{2}         \and
         A.~Strza{\l}kowski\inst{4}     \and
	 A. T\"aschner\inst{2}         \and
         M.~Wolke\inst{1}               \and
         P.~W{\"u}stner\inst{5}         \and
         W.~Zipper\inst{6}               
}                     
\institute{IKP, Forschungszentrum J\"{u}lich, D-52425 J\"{u}lich, Germany
     \and IKP, Westf\"{a}lische Wilhelms--Universit\"{a}t, D-48149 M\"{u}nster, Germany
     \and Institute of Nuclear Physics, PL-31-342 Cracow, Poland 
     \and M.~Smoluchowski Institute of Physics, Jagellonian University, PL-30-059 Cracow, Poland
     \and ZEL,  Forschungszentrum J\"{u}lich, D-52425 J\"{u}lich,  Germany
     \and Institute of Physics, University of Silesia, PL-40-007 Katowice, Poland}
\date{Received: date / Revised version: date}
\abstract{
At the internal facility COSY-11 a first measurement of the reaction \reaktion near the production threshold has been performed. Results for the analysing power will be presented and a comparison with one meson exchange models will be discussed.
\PACS{
      {12.40.Vv}{Vector-meson dominance} \and
      {13.60.Le}{Meson production} \and
      {13.88.+e}{Polarization in interactions and scattering} \and
      {24.70.+s}{Polarization phenomena in reactions} \and
      {24.80.+y}{Nuclear tests of fundamental interactions and symmetries} \and
      {25.10.+s}{Nuclear reactions involving few-nucleon systems}
     } 
} 
\begin{document}
\maketitle
\section{Introduction}
\label{intro}
Triggered by an extensive database on $\eta$ meson production in $NN$ collisions through measurements of total \cite{bergdolt:93,chiavassa:94,calen:96,calen:97,hibou:98,smyrski:00} as well as differential cross sections \cite{calen:99,tatischeff:00,moskal:01-2,abdelbary:02,moskal:02-2} lots of theoretical investigations have been performed in this field of physics over the last years. Several one meson exchange models -- differing mainly in the assumptions for the production mechanism -- reproduce the existing data quite well so that additional measurements are needed. Polarisation observables -- analysing powers or spin correlation coefficients -- present a powerful tool because they are sensitive to the influence of higher partial waves.

\section{Experiment}
\label{sec:1}
\begin{figure}[ht]
\centerline{\epsfig{file=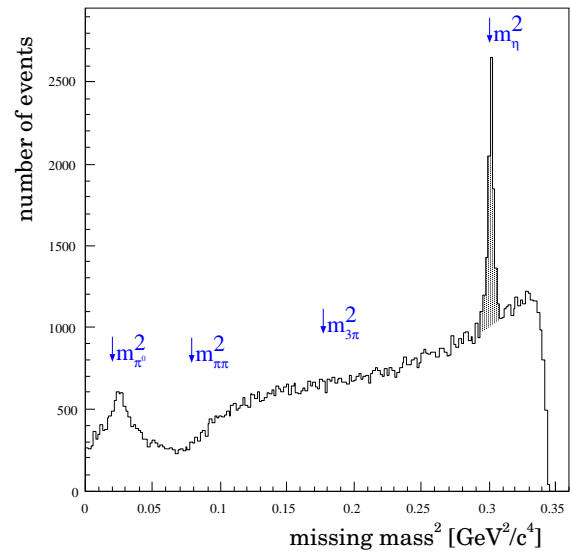,width=0.85\columnwidth}}
\caption{Missing mass spectrum of events with two identified protons in the exit channel.}
\label{fig:1}       
\end{figure}

The measurement was performed at the internal facility COSY-11 \cite{brauksiepe:96,moskal:01} at the COler SYnchrotron COSY \cite{maier:97nim} in J\"ulich with a beam momentum of p$_{beam}=2.096\,$GeV/c corresponding to an excess energy of $Q=40\,$MeV. The reaction takes place in a cluster target \cite{dombrowski:97} mounted in front of a COSY dipole magnet. A set of drift chambers and a time-of-flight measurement with scintillation detectors allows for a four momentum determination of positively charged ejectiles. An identification of the unregistered meson succeeds with the missing mass method (see Figure~\ref{fig:1}). 

Besides a clear signal from the $\eta$ meson there is a broad background due to multi pion reactions. The rising shape of this background results from the acceptance behaviour of the detection system. A subtraction of the background with a polynomial fit allows the determination of the number of $\eta$-events. An efficiency correction for the polar and azimuthal angle of the relative proton-proton momentum in the pp-rest system was applied. This is necessary for an extraction of interference terms from contributing partial waves. The detailed analysing procedure is described in \cite{winter:02-2,winter:02-1-en}. 

\section{Results}
\label{sec:2}
\begin{figure}[ht]
\centerline{\epsfig{file=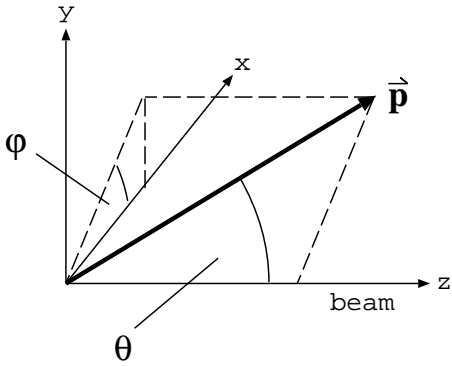,scale=1}}
\caption{Definition of the angles. $\theta$ is defined as the angle between momentum vector and the z-axis, $\varphi$ between the x-axis and the projection of $\vec{p}$\, onto the x-y-plane.}
\label{fig:2}       
\end{figure}

For later purposes, Figure \ref{fig:2} depicts the definition of the used polar- $(\theta)$ and azimuthal angle $(\varphi)$. The indices $p$ and $q$ refer to the pp rest-system and the $\eta$ meson in the CMS, respectively. The angle $\theta_p$ is chosen such, that $0\leq\theta_p\leq\pi/2$. This choice guarantees that all observables are invariant under the transformation $\vec{p} \rightarrow -\vec{p}$ as required by the identity of the two protons in the final state.

The determination of the analysing power needs the knowledge of the averaged beam polarisation P$_{\uparrow,\downarrow}$ for the cycles with spin up and down, respectively, the relative time-integrated luminosity $\mathcal{L}_{rel} :=\int \mathcal{L}_\downarrow\,dt_\downarrow\, /\, \int \mathcal{L}_\uparrow\,dt_\uparrow$ and the number of events N$_{\uparrow,\downarrow}$. 

Via a simultaneous measurement of the $\vec{\mbox{p}}$p-elastic scattering at the internal experiment EDDA \cite{albers:97,altmeier:00} the beam polarisation was determined for two time blocks (see table~\ref{tab:1}). The slight increase between the two blocks was caused by improved tuning of the beam with respect to polarisation.

\begin{table}
\caption{Averaged beam polarisation obtained with the simultaneous measurement at the EDDA experiment}
\label{tab:1}       
\centerline{\begin{tabular}{lcc}
\hline\noalign{\smallskip}
& time block 1 & time block 2 \\
\noalign{\smallskip}\hline\noalign{\smallskip}
P$_\uparrow$ & $0.381\pm 0.007$ & $0.497\pm 0.006$ \\
P$_\downarrow$ & $-0.498\pm 0.007$ & $-0.572\pm 0.007$ \\
\noalign{\smallskip}\hline
\end{tabular}}
\end{table}

The relative luminosity was extracted from the comparison of the measured angular distribution of the elastically scattered protons with the distribution known from literature \cite{albers:97,altmeier:00}.

\begin{table}[ht]
\caption{Time-integrated relative luminosity  $\mathcal{L}_{rel}$}
\label{tab:2}       
\centerline{\begin{tabular}{lcc}
\hline\noalign{\smallskip}
& time block 1 & time block 2 \\
\noalign{\smallskip}\hline\noalign{\smallskip}
$\mathcal{L}_{rel}$ & $1.004\pm 0.004 {+ 0.002 \atop - 0.002}$ & $0.949 \pm 0.004 {+ 0.001 \atop - 0.001}$ \\
\noalign{\smallskip}\hline
\end{tabular}}
\end{table}

In order to account for the detection efficiency the data has been corrected on an event-by-event basis by the weighting factors determined via Monte-Carlo simulations \cite{winter:02-2,winter:02-1-en}. Using a GEANT-3 code, $10^7$ events were generated and for each event a detection system response was calculated. The simulated data sample was analysed with the same programme which is used for the analysis of the experimental data. 

The determined averaged values $\bar{A}_y(\cos\theta_q^*)$ of the ana\-lysing power for the \reaktion reaction are presented in table \ref{tab:3} as a function of the center of mass polar angle $\cos\theta_q^*$ of the $\eta$ meson. Explicitely, $\bar{A}_y(\cos\theta_q^*)$ is defined via
\begin{eqnarray}
\lefteqn{\bar{A}_y(\cos\theta_q^*) :=}\\
&&  \iiint \frac{d^2\sigma}{d\Omega_p d\Omega_q}(\xi)\,A_y(\xi)\, d\cos\theta_p d\varphi_p d\varphi^*_q \,
 /\, \frac{d\sigma}{d\cos\theta^*_q}\,,
\label{eq:1}
\end{eqnarray}
where $\xi$ denotes the set of the five variables which kinematically completely describe the exit channel, namely $(\theta_p,\ \varphi_p,\ \theta^*_q,\ \varphi^*_q,\ E_{pp})$. The kinetic energy $E_{pp}$ of the two final protons in their CM system is given by $E_{pp} = \sqrt{s_{12}} - 2m_p$ with $ \sqrt{s_{12}} = 2\sqrt{\vec{p}^2 + m_p^2}$ as the energy in the pp subsystem. $\theta_q^*$ and $\varphi^*_q$ denote the angles of the $\eta$ meson in the CMS.

\begin{table}[ht]
\caption{Analysing power as a function of the emission angle $\theta^*_q$ of the $\eta$ meson in the CMS.}
\label{tab:3}       
\centerline{\begin{tabular}{ccc}
\hline\noalign{\smallskip}
$\boldsymbol{\cos\theta^*_q}$ && $\boldsymbol{\bar{A_y}(\cos\theta_q^*)}$ \\
\noalign{\smallskip}\hline\noalign{\smallskip}
-0.75\,$\pm$\,0.25 && 0.19\,$\pm$\,0.21 \\
-0.25\,$\pm$\,0.25 && -0.02\,$\pm$\,0.09 \\
 0.25\,$\pm$\,0.25 && 0.05\,$\pm$\,0.06 \\
 0.75\,$\pm$\,0.25 && -0.05\,$\pm$\,0.06 \\
\noalign{\smallskip}\hline
\end{tabular}}
\end{table}

For the derivation of these values for the analysing power $\bar{A}_y(\cos\theta_q^*)$,  we used an ansatz for the spin-dependent cross section as already applied in case of $\vec{p}\vec{p}\rightarrow\,pp\pi^0$ reaction \cite{meyer:01}. With the inclusion of the efficiency correction, an integration gives (details can be reviewed in \cite{winter:02-2}):
\begin{eqnarray}
\lefteqn{\iint \frac{d^2\sigma}{d\Omega_p d\Omega_q}(\xi)\,A_y(\xi)\, d\cos\theta_p\,d\varphi_p =} \nonumber \\
& & 2\pi\left(G_1^{y0}\sin\theta^*_q + (H_1^{y0}+I^{y0})\sin 2\theta^*_q\right) \cos\varphi^*_q,
\label{eq:2}
\end{eqnarray}
where $G_1^{y0},\ H_1^{y0} $ and $I^{y0}$ described in \cite{meyer:01} correspond to the explicit interference terms (PsPp), (Pp)$^2$ and (SsSd) of the partial wave amplitudes. Here, the relative angular momentum of the two outgoing protons in their rest system is denoted by capital letters $l_p=S,P,D\ldots$, the one of the $\eta$ meson in the CMS by small letters $l_q=s,p,d\ldots$. 

With the equations \eqref{eq:1} and \eqref{eq:2}, it can be shown that an extraction of the interference terms $G_1^{y0}$ and $H_1^{y0}+I^{y0}$ is possible from the experimental data via
\begin{eqnarray*}
G_1^{y0} & = & \frac{1}{\pi^2} \sum_{\cos\theta^*_q} \frac{d\sigma}{d\cos\theta^*_q} \bar{A}_y\cdot \Delta\cos\theta^*_q\\
H_1^{y0}+I^{y0} & = & \frac{2}{\pi^2} \sum_{\cos\theta^*_q} \frac{d\sigma}{d\cos\theta^*_q} \bar{A}_y \cos\theta^*_q\cdot \Delta\cos\theta^*_q\,.
\end{eqnarray*}
and results in 
\[
G_1^{y0}= (0.003\pm 0.004)\,\mu\mbox{b}
\]
and 
\[
H_1^{y0}+I^{y0}=(-0.005\pm 0.005)\,\mu\mbox{b}\,. 
\]

\section{Comparison with theory}
\begin{figure}[ht]
\centerline{\epsfig{file=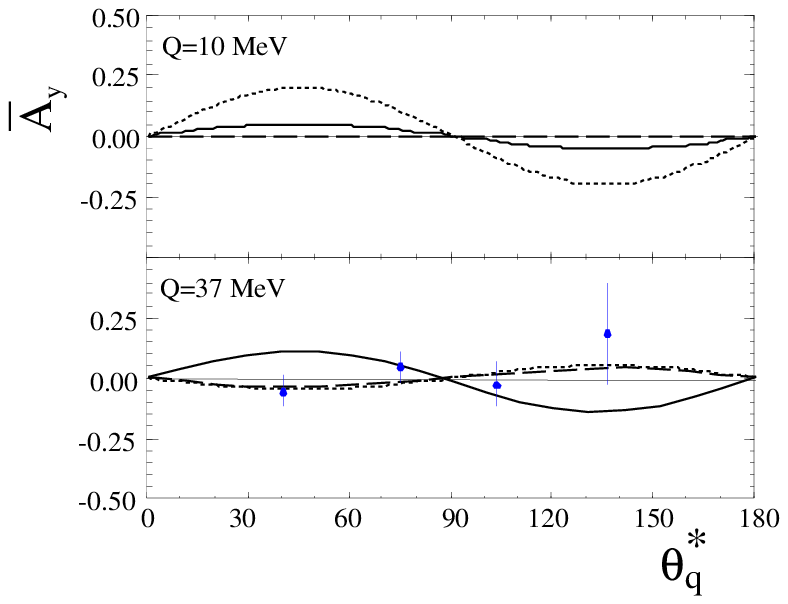,scale=1.05}}
\caption{Analysing power for the reaction \reaktion in dependence on $\theta^*_q$ for the two excess energies $Q=10\,$MeV and 37\,MeV.}
\label{fig:3}
\end{figure}

Figure \ref{fig:3} shows a comparison of the data (triangles) with two model predictions taken from \cite{faeldt:01} (dotted line) and \cite{nakayama:02} (solid and dashed lines) for $Q=10\,$MeV and 37\,MeV.

While F\"aldt and Wilkin \cite{faeldt:01} conclude a dominant $\rho$ meson exchange for the underlying production mechanism, the authors of reference \cite{nakayama:02} find a dominance of $\pi$ and $\eta$-exchange (solid line). The dashed curve represents a reduction of their full model to a vector dominance model with an exclusion of $\pi$ and $\eta$-exchange. It seems that the data favours slightly the vector dominance exchange models. 

The observable structure of the experimental values show a slight deviation from the $\sin\theta_q^*\cos\theta^*_q$-dependence of both models. This deviation indicates a non-vanishing value of $G_1^{y0}$. As this corresponds to the (PpPs)-term, an influence of the P-wave must be suspected but right now the experimental result for $G_1^{y0}$ is compatible with zero. A non-zero $G_1^{y0}$ would imply that $H_1^{y0}$ -- describing the (Pp)$^2$ interference -- should have a non negligible contribution, too. For further detailed studies the data are not yet precise enough to disentangle the sum of $H_1^{y0}$ and $I^{y0}$. Therefore, new data at $Q=37$\,MeV has been taken in September 2002 in order to reduce the error bars by more than a factor of 2. The data analysis is currently in progress but it is already obvious that the much higher statistics and polarisation of $P\approx 80\%$ will enable to achieve a significant increase in precision. 

Furthermore, additional measurements are scheduled in the first half of 2003 at $Q=2,\ 10$ and $25\,$MeV for tests on the predictions of the different models for the energy dependence of $A_y(\theta_q^*)$.

\section{Acknowledgements}
This work was partly supported by the European Community -- Access to Research Infrastructure action of the Improving Human Potential Programme.

\bibliographystyle{proceedings}
\bibliography{abbrev,polarisation,general}

\end{document}